\documentclass[final,5p,times,twocolumn]{elsarticle}
\pdfoutput=1

\usepackage{amsmath,amssymb,amsfonts,bm,mathtools}
\usepackage{graphicx}
\usepackage{booktabs}
\usepackage{multirow}
\usepackage{slashed}
\usepackage{hyperref}
\usepackage{microtype}
\usepackage{dblfloatfix}

\newcommand{\dd}{\mathrm{d}}

\biboptions{numbers,sort&compress}
\journal{Physics Letters B}

\begin{document}
\begin{frontmatter}

\title{Quantum-Kinetic Leptogenesis and Gravitational Waves from Seesaw-Assisted Domain-Wall Dynamics}

\author{Gayatri Ghosh}
\address{Department of Physics, Cachar College, Silchar, Assam, India}

\begin{abstract}
We investigate the connection between resonant leptogenesis and a
primordial stochastic gravitational-wave background in a minimal
two-right-handed-neutrino type-I seesaw with a real singlet scalar.
The scalar sector admits a $\mathbb Z_2$-symmetric tree-level potential,
while the tiny right-handed-neutrino coupling to the $\mathbb Z_2$-odd
scalar generates a radiative vacuum-energy bias that causes the
domain-wall network to annihilate. For quasi-degenerate heavy neutrinos,
we describe the heavy-neutrino system with a density-matrix kinetic
framework and use the ratio $r_q=\Delta M/\Gamma_{N_1}$ as an organizing
variable for the separated, resonant and coherent regimes. The heavy-neutrino parameters that control the physical mass splitting also enter the radiatively generated domain-wall bias, providing a
model-dependent link between the quantum-kinetic baryon asymmetry and
the gravitational-wave peak frequency and amplitude. We emphasize that the gravitational-
wave signal is not a direct measurement of an individual low-energy
neutrino parameter; rather, it provides a model-dependent consistency
relation among neutrino data, resonant leptogenesis, domain-wall
annihilation and the stochastic gravitational-wave background.
\end{abstract}

\end{frontmatter}

\section{Introduction}

The origin of neutrino masses and the baryon asymmetry of the Universe
are among the clearest indications of physics beyond the Standard Model.
The type-I seesaw mechanism provides a minimal and economical framework
in which light-neutrino masses arise from heavy right-handed Majorana
neutrinos~\cite{Minkowski1977,GellMann1979,Yanagida1980,SchechterValle1980,Weinberg1979}.
The same heavy states can generate a lepton asymmetry through
CP-violating out-of-equilibrium dynamics, which is subsequently converted
into a baryon asymmetry through electroweak sphalerons~\cite{FukugitaYanagida1986}.
While hierarchical thermal leptogenesis generally favors a high seesaw
scale, a quasi-degenerate heavy-neutrino spectrum can resonantly enhance
the CP asymmetry and allow successful leptogenesis over a much broader
range of masses~\cite{Pilaftsis1997,PilaftsisUnderwood2004}.

An important question is whether such a weakly coupled neutrino sector
can be probed indirectly. Primordial gravitational waves provide a
promising possibility because the early Universe can contain topological
defects and phase transitions whose characteristic scales are related
to particle-physics parameters~\cite{Caprini2019}. In particular,
domain walls produced by the spontaneous breaking of a discrete symmetry
can annihilate in the presence of a small bias and generate a stochastic
gravitational-wave background~\cite{Saikawa2017,Hiramatsu2013}.

A particularly interesting realization combines the seesaw sector with
a scalar field responsible for the formation and subsequent annihilation
of domain walls. In such a framework, the scalar vacuum expectation value
can contribute to the right-handed-neutrino mass matrix, while the same
scalar--neutrino interactions can generate a radiative vacuum-energy
bias between the two degenerate vacua. The resulting domain-wall
gravitational-wave signal can therefore retain information about the
heavy-neutrino sector. Recent seesaw-assisted and generic $Z_2$ domain-wall studies have established
this radiative-bias mechanism and its gravitational-wave implications~\cite{KitajimaLeeMuraiTakahashiYin2024,BanerjeeYajnik2024}.
Related connections between right-handed neutrinos, leptogenesis and
primordial gravitational waves have also been explored through
cosmological defects, cosmic strings and supercooled phase transitions
~\cite{Dror2020,Athron2026,Datta2026}.

The present work addresses the more specific question: can the
\emph{quantum-coherent dynamics} of a quasi-degenerate right-handed-neutrino
system be quantitatively mapped onto the gravitational-wave parameter
space of a seesaw-assisted domain-wall scenario. When the heavy-neutrino
mass splitting becomes comparable to the interaction-induced damping
rate, the two heavy-neutrino states cannot in general be treated as
independent classical populations. The off-diagonal components of their
density matrix encode the coherent oscillations and are essential for a
consistent description of resonant leptogenesis~\cite{Dev2014,Jukkala2021}.

We therefore formulate the heavy-neutrino evolution in a density-matrix
quantum-kinetic framework and introduce the dimensionless parameter
\begin{equation}
r_q \equiv \frac{\Delta M}{\Gamma_{N_1}},
\end{equation}
where $\Delta M=M_{N_2}-M_{N_1}$ is the heavy-neutrino mass splitting
and $\Gamma_{N_1}$ is the decay width of the lighter state. The three
regimes
\begin{equation}
r_q\gg1,\qquad r_q\simeq1,\qquad r_q\ll1
\end{equation}
correspond respectively to effectively separated states,
resonant-overlapping dynamics and strongly coherent evolution.
The use of density-matrix quantum kinetics in resonant leptogenesis is
well established~\cite{Dev2014,Jukkala2021}; the new aspect here is to
propagate this quantum-kinetic structure into the domain-wall
gravitational-wave sector.

The heavy-neutrino parameters play a dual role in the leptogenesis
and domain-wall sectors. In the leptogenesis sector, the physical
mass splitting
\begin{equation}
\Delta M
=
(M_2-M_1)-(y_2-y_1)v_\phi
\label{eq:physical-splitting}
\end{equation}
controls the degree of heavy-neutrino overlap through
\begin{equation}
\Delta M
\longrightarrow
\frac{\Delta M}{\Gamma_{N_1}}
\longrightarrow
Y_B^{\rm QKE}.
\label{eq:QKE-chain}
\end{equation}
In the domain-wall sector, however, the radiative vacuum-energy bias
is determined by the scalar effective potential evaluated at the two
vacua and depends, in general, on the individual heavy-neutrino
parameters $M_i$ and $y_i$, rather than on $\Delta M$ alone. The
microscopic dependence is therefore more appropriately represented as
\begin{equation}
\{M_1,M_2,y_1,y_2,v_\phi,\lambda_\phi\}
\longrightarrow
\{\Delta M,\Delta V_{\rm bias}\},
\label{eq:microscopic-map-intro}
\end{equation}
followed by
\begin{equation}
\Delta V_{\rm bias}
\longrightarrow
T_{\rm ann}
\longrightarrow
\left(
f_{\rm peak},
\Omega_{\rm GW}^{\rm peak}h^2
\right).
\label{eq:DW-map-intro}
\end{equation}
Thus, the two sectors are correlated through their common dependence
on the underlying microscopic parameters, but $\Delta M$ does not
uniquely determine $\Delta V_{\rm bias}$. The resulting connection is
therefore a model-dependent multi-parameter correlation rather than a
one-to-one mapping between the heavy-neutrino mass splitting and the
gravitational-wave observables.

Our novelty therefore does not lie in the generic association of
seesaw neutrinos, domain walls and gravitational waves, which has
already been established in the literature~\cite{Dror2020,KitajimaLeeMuraiTakahashiYin2024,BanerjeeYajnik2024}.
Rather, we identify the quantum-coherence ratio
$\Delta M/\Gamma_{N_1}$ as the microscopic coordinate that organizes
the leptogenesis regimes, while the domain-wall gravitational-wave
signal is obtained from the same underlying heavy-neutrino and scalar
parameters through the radiatively generated vacuum-energy bias.
The resulting relation is therefore a correlated projection of the
microscopic parameter space rather than a direct one-parameter map
from $\Delta M/\Gamma_{N_1}$ to the gravitational-wave observables. In particular, we compare the
conventional Boltzmann treatment with the density-matrix evolution and
investigate whether the quantum-resonant region
$\Delta M\sim\Gamma_{N_1}$ produces a distinguishable GW prediction.

The numerical analysis simultaneously propagates the same microscopic parameter point through the kinetic and domain-wall calculations.
The parameter scan is organized in the
$(M_{N_1},v_\phi)$, $(M_{N_1},\Delta M)$ and
$(\Delta M/\Gamma_{N_1},v_\phi)$ planes. The phenomenologically relevant
region is selected by imposing the observed baryon asymmetry together
with the consistency conditions on domain-wall annihilation and the
corresponding cosmological gravitational-wave bounds. The resulting
scan therefore provides a direct numerical test of whether the
quantum-coherent RHN regime can leave an observable imprint on the
primordial stochastic gravitational-wave background. 

\section{Minimal seesaw setup}
\label{sec:model}

We consider a minimal type-I seesaw framework containing two right-handed
Majorana neutrinos, $N_{1,2}$, and a real singlet scalar $\phi$. The
additional scalar is responsible for the spontaneous breaking of a
discrete $\mathbb{Z}_2$ symmetry and hence for the formation of a domain-wall
network in the early Universe. At the same time, its coupling to the
right-handed neutrinos modifies their mass matrix and provides the
microscopic origin of the heavy-neutrino mass splitting relevant for
resonant leptogenesis. This realizes the basic seesaw--domain-wall
framework studied recently in Ref.~\cite{KitajimaLeeMuraiTakahashiYin2024}, while our
subsequent analysis focuses on the quantum-kinetic evolution of the
quasi-degenerate $N_{1,2}$ system. The general role of right-handed
neutrinos in minimal and low-scale seesaw constructions is reviewed,
for example, in Refs.~\cite{King2025}.

The relevant fermionic interactions are
\begin{equation}
-\mathcal{L}_{N}
\supset
h_{\alpha i}\,
\overline{L_\alpha}\widetilde{H}N_i
+
\frac{1}{2}M_i\,
\overline{N_i^c}N_i
+
\frac{1}{2}y_i\phi\,
\overline{N_i^c}N_i
+\mathrm{h.c.},
\label{eq:lagrangian}
\end{equation}
where $\alpha=e,\mu,\tau$ and $i=1,2$, $\widetilde H=i\sigma_2H^*$,
$h_{\alpha i}$ denotes the neutrino Yukawa coupling matrix and $y_i$
the coupling of the singlet scalar to the right-handed neutrinos.

We assign
\begin{equation}
\phi\rightarrow-\phi,
\qquad
N_i\rightarrow N_i,
\end{equation}
while all Standard Model fields are taken to be $\mathbb{Z}_2$ even.
The scalar potential is chosen to be $\mathbb{Z}_2$ symmetric at tree
level. The $\phi\,\overline{N_i^c}N_i$ interaction, however, is odd under
this transformation for the above charge assignment and therefore
provides the explicit symmetry-breaking interaction responsible for
lifting the degeneracy of the two vacua once quantum corrections are
included. This distinction is important: the tree-level scalar
potential possesses the degenerate $\mathbb{Z}_2$ vacua, whereas the
right-handed-neutrino sector generates a calculable vacuum-energy
difference between them. Such radiatively induced biases provide a
well-motivated mechanism for destabilizing otherwise cosmologically
dangerous domain walls~\cite{Saikawa2017,Hiramatsu2013,Notari2025,KitajimaLeeMuraiTakahashiYin2024,BanerjeeYajnik2024}.

\begin{figure}[t]
    \centering
    \includegraphics[width=\columnwidth]{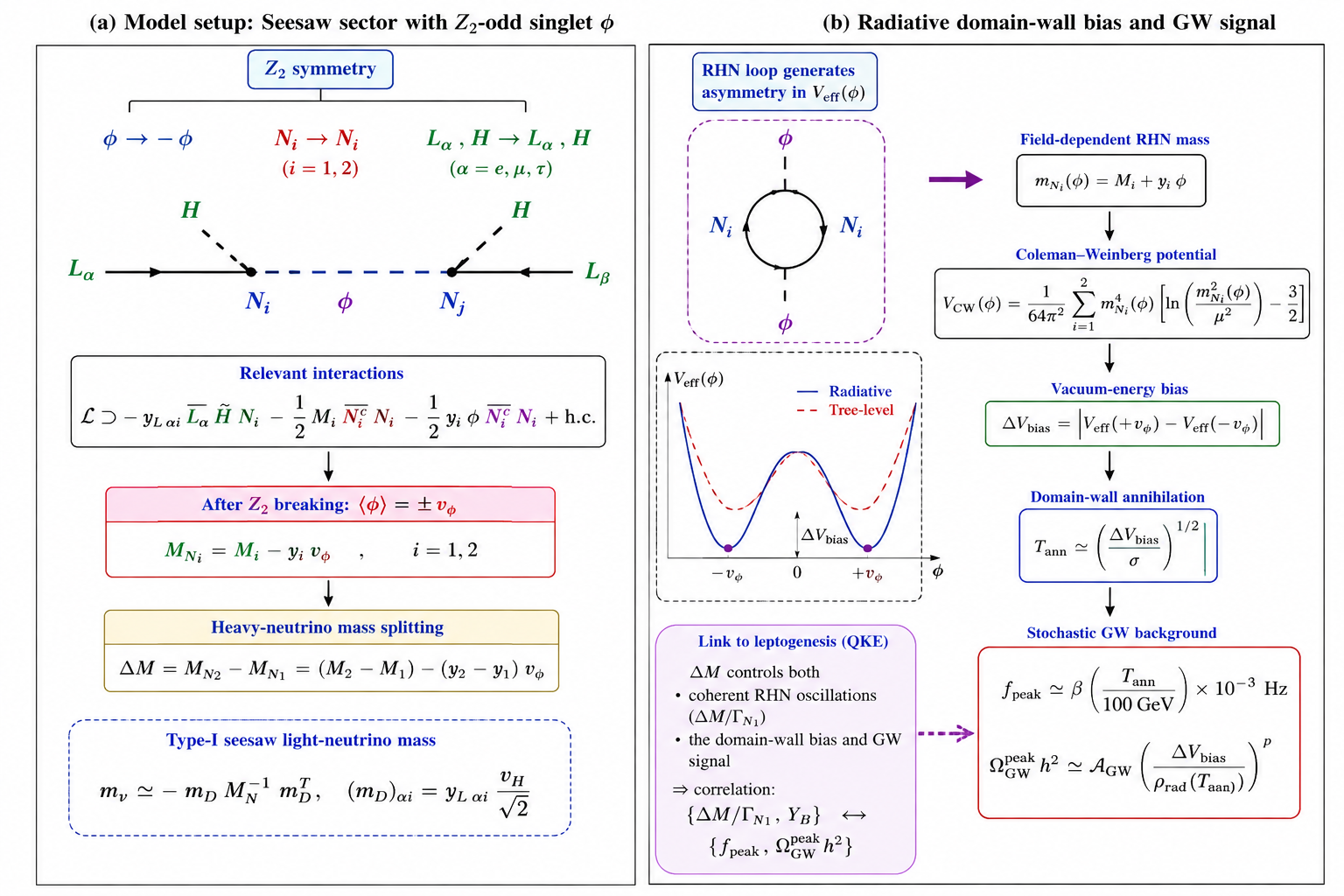}
    \caption{
    Schematic overview of the minimal $Z_2$-odd singlet plus two-right-handed-neutrino
    seesaw framework considered in this work. The upper panel summarizes the
    particle content and interactions and shows how $\langle\phi\rangle=\pm v_\phi$
    modifies the heavy-neutrino masses. The lower panel illustrates the radiative
    generation of the vacuum-energy bias by the right-handed-neutrino sector,
    which drives domain-wall annihilation and determines the resulting
    gravitational-wave signal, following the established domain-wall and
    seesaw--GW literature~\cite{Hiramatsu2013,Saikawa2017,KitajimaLeeMuraiTakahashiYin2024,BanerjeeYajnik2024,
    Caprini2019,Dror2020}. The same heavy-neutrino mass splitting $\Delta M$
    controls the quantum-kinetic dynamics through
    $\Delta M/\Gamma_{N_1}$, thereby linking resonant leptogenesis to the
    domain-wall gravitational-wave spectrum~\cite{Pilaftsis1997,PilaftsisUnderwood2004,
    GarnyKartavtsevHohenegger2013,Dev2014,Jukkala2021}.
    }
    \label{fig:model_overview}
\end{figure}

The particle content and the sequence
The relevant microscopic sequence is
$\phi$ breaking $\rightarrow(M_{N_i},\Delta M)\rightarrow$ quantum
kinetics and radiative bias $\rightarrow
(f_{\rm peak},\Omega_{\rm GW}^{\rm peak}h^2)$; it is summarized in
Fig.~\ref{fig:model_overview}.
are summarized schematically in Fig.~\ref{fig:model_overview}. The figure
is intended as a model overview rather than as a numerical result.

\subsection{Scalar symmetry breaking and the heavy-neutrino spectrum}
\label{subsec:scalar_seesaw}

Neglecting the Higgs portal interaction in the minimal setup, the
tree-level scalar potential is
\begin{equation}
V(\phi)
=
-\frac{\mu_\phi^2}{2}\phi^2
+\frac{\lambda_\phi}{4}\phi^4 ,
\label{eq:scalar-potential}
\end{equation}
with $\lambda_\phi>0$. The $\mathbb{Z}_2$ symmetry is spontaneously broken
when $\mu_\phi^2>0$, giving
\begin{equation}
\langle\phi\rangle=\pm v_\phi,
\qquad
v_\phi=\frac{\mu_\phi}{\sqrt{\lambda_\phi}}.
\label{eq:vphi}
\end{equation}
The two degenerate minima correspond to distinct $\mathbb{Z}_2$ vacua and
lead to the formation of a domain-wall network after the phase transition.
For the quartic potential in Eq.~\eqref{eq:scalar-potential}, the
zero-temperature kink solution gives
\begin{equation}
\sigma_{\rm DW}=
\frac{2\sqrt{2\lambda_\phi}}{3}\,v_\phi^3 .
\label{eq:sigma_scaling}
\end{equation}
The cosmological evolution of such walls, including their annihilation
and the resulting stochastic gravitational-wave spectrum, has been
studied extensively~\cite{Saikawa2017,Hiramatsu2013,Caprini2019}. Recent
lattice simulations further indicate that the detailed GW spectrum can
depend on the dynamics of the annihilation stage and need not be captured
completely by the simplest analytic estimate~\cite{Notari2025}.

Choosing the vacuum $\langle\phi\rangle=-v_\phi$, the heavy-neutrino mass
matrix becomes
\begin{equation}
M_N=M-yv_\phi ,
\label{eq:MN}
\end{equation}
where $M$ and $y$ denote the corresponding matrices in the heavy-neutrino
flavour space. For the diagonal benchmark adopted in the numerical
analysis,
\begin{equation}
M_{N_i}=M_i-y_i v_\phi .
\label{eq:MN-diagonal}
\end{equation}
Consequently, the physical mass splitting is
\begin{equation}
\Delta M
\equiv
M_{N_2}-M_{N_1}
=
(M_2-M_1)-(y_2-y_1)v_\phi .
\label{eq:mass-splitting}
\end{equation}
In the quasi-degenerate regime, $\Delta M\ll M_{N_1}$, this splitting
becomes the key microscopic parameter controlling the overlap and
coherent evolution of the two heavy-neutrino states. In particular, the
relevant quantum-kinetic parameter introduced below is
\begin{equation}
r_q\equiv\frac{\Delta M}{\Gamma_{N_1}},
\label{eq:rq}
\end{equation}
where $\Gamma_{N_1}$ is the interaction-induced decay width of $N_1$.
The limits $r_q\gg1$, $r_q\sim1$ and $r_q\ll1$ correspond, respectively,
to effectively separated states, resonant overlap and strong quantum
coherence. Density-matrix treatments are required when the mass
splitting becomes comparable to the interaction rate, as emphasized in
the resonant-leptogenesis literature~\cite{Dev2014,Jukkala2021}.

\subsection{Light-neutrino masses}
\label{subsec:light_neutrino}

After electroweak symmetry breaking,
\begin{equation}
H=
\frac{1}{\sqrt{2}}
\begin{pmatrix}
0\\
v+h
\end{pmatrix},
\qquad
v=246~{\rm GeV},
\end{equation}
and the Dirac mass matrix is
\begin{equation}
(M_D)_{\alpha i}
=
\frac{v}{\sqrt{2}}h_{\alpha i}.
\label{eq:MD}
\end{equation}
In the seesaw limit,
\begin{equation}
M_D\ll M_N,
\end{equation}
the effective light-neutrino mass matrix is
\begin{equation}
M_\nu
\simeq
-M_D M_N^{-1}M_D^T .
\label{eq:seesaw}
\end{equation}
This is the standard type-I seesaw relation~\cite{Minkowski1977,Yanagida1980,
GellMann1979,SchechterValle1980}. Since only two heavy Majorana neutrinos
are introduced, the minimal construction predicts one vanishing
light-neutrino mass at tree level,
\begin{equation}
m_{\rm lightest}\simeq0,
\end{equation}
up to possible radiative or higher-dimensional corrections. The remaining
light-neutrino masses and mixing parameters are fixed by neutrino
oscillation data, while the high-energy freedom can conveniently be
parameterized using the Casas--Ibarra construction~\cite{CasasIbarra2001}. The connection with the low-energy neutrino sector and mixing structure is complementary to earlier seesaw and flavour analyses~\cite{King2000,Abada2006,Antusch2009}.
This separation between the measured low-energy neutrino sector and the
high-energy seesaw parameters is particularly useful when imposing
oscillation constraints simultaneously with the leptogenesis and
gravitational-wave requirements.

The heavy-neutrino spectrum therefore enters the cosmological dynamics
through two complementary combinations. The overall scale $M_{N_1}$
controls the seesaw and heavy-neutrino interaction rates, whereas the
splitting $\Delta M$ controls the degree of resonant overlap. At the same
time, $\Delta M$ enters the scalar-induced vacuum-energy bias and hence
the domain-wall annihilation history. This dual role is the central
structural feature exploited in the subsequent quantum-kinetic and
gravitational-wave analysis.

\subsection{Light-neutrino constraints and Casas--Ibarra parametrization}
\label{subsec:CI}

The parameters of the neutrino Yukawa sector are constrained by the
measured light-neutrino masses and leptonic mixing parameters. We impose
these constraints using the Casas--Ibarra parametrization
\cite{CasasIbarra2001},
\begin{equation}
 h
 =
 \frac{\sqrt{2}}{v}\,
 U_{\rm PMNS}\,
 \sqrt{m_\nu^{\rm diag}}\,
 R^\dagger\,
 \sqrt{M_N},
\label{eq:CI}
\end{equation}
where
\begin{equation}
 m_\nu^{\rm diag}
 =
 {\rm diag}(m_1,m_2,m_3),
\end{equation}
$U_{\rm PMNS}$ denotes the leptonic mixing matrix and $R$ is a complex
orthogonal matrix satisfying
\begin{equation}
 RR^T=\mathbb{I}_{2\times2}.
\end{equation}
Because the present framework contains only two right-handed neutrinos,
the tree-level light-neutrino mass matrix has rank two and one light
neutrino is massless. We therefore consider both normal ordering (NO),
for which
\begin{equation}
m_1=0,
\end{equation}
and inverted ordering (IO), for which
\begin{equation}
m_3=0.
\end{equation}

For two right-handed neutrinos, the complex orthogonal matrix can be
parameterized by a single complex angle,
\begin{equation}
z=x+iy,
\end{equation}
where $x$ and $y$ are real. For normal ordering we use
\begin{equation}
 R_{\rm NO}
 =
 \begin{pmatrix}
 0 & \cos z & \sin z\\
 0 & -\sin z & \cos z
 \end{pmatrix},
\label{eq:RNO}
\end{equation}
with the corresponding permutation for inverted ordering. The Casas--Ibarra
construction allows the measured low-energy neutrino parameters to be
kept fixed while varying the high-energy seesaw parameters. In particular,
the heavy-neutrino scale, mass splitting and the complex angle $z$
determine the decay and interaction rates entering the leptogenesis
calculation, whereas the low-energy masses and mixing angles are fixed
by neutrino-oscillation data.

This separation is particularly useful for the present analysis. The
same heavy-neutrino parameters that determine the Yukawa combination
$h^\dagger h$ and hence the interaction rate $\Gamma_{N_1}$ also enter
the mass splitting
\begin{equation}
\Delta M=M_{N_2}-M_{N_1}.
\end{equation}
The ratio
\begin{equation}
r_q\equiv
\frac{\Delta M}{\Gamma_{N_1}}
\end{equation}
therefore provides a convenient dimensionless measure of the quantum
regime of the quasi-degenerate heavy-neutrino system. The limits
$r_q\gg1$, $r_q\sim1$ and $r_q\ll1$ correspond respectively to
effectively separated states, resonant overlap and strong quantum
coherence. A density-matrix treatment becomes essential when the mass
splitting is comparable to the interaction rate
\cite{Dev2014,Jukkala2021}.

\section{Domain walls and their gravitational-wave signal}
\label{sec:dw}

The spontaneous breaking
\begin{equation}
\mathbb{Z}_2\longrightarrow\mathbb{I}
\end{equation}
produces two degenerate vacua,
\begin{equation}
\phi=\pm v_\phi,
\end{equation}
and consequently a network of domain walls in the early Universe.
In the absence of a bias, a stable domain-wall network can eventually
dominate the energy density and is therefore cosmologically
unacceptable. A small vacuum-energy difference between the two vacua
removes the degeneracy and causes the walls to collapse
\cite{Hiramatsu2013,Saikawa2017}.

In the present framework, the coupling of the $Z_2$-odd scalar to the
right-handed neutrinos provides a microscopic source of this bias through
the radiative effective potential. This is an important structural
feature of the model because the same RHN parameters enter both the
seesaw/leptogenesis sector and the domain-wall dynamics. A closely
related seesaw-assisted domain-wall mechanism and its gravitational-wave
phenomenology were recently studied in Ref.~\cite{KitajimaLeeMuraiTakahashiYin2024}. The
purpose here is not to re-establish the existence of this generic
seesaw--domain-wall--GW connection, but to investigate its extension
when the quasi-degenerate RHN system is treated quantum mechanically.

The zero-temperature one-loop contribution from the heavy Majorana
neutrinos to the scalar effective potential is
\begin{equation}
 V_{\rm CW}(\phi)
 =
 -\frac{1}{32\pi^2}
 \sum_i m_{N_i}^4(\phi)
 \left[
 \ln\left(
 \frac{m_{N_i}^2(\phi)}{\mu_R^2}
 \right)
 -\frac{3}{2}
 \right],
\label{eq:CW}
\end{equation}
where
\begin{equation}
m_{N_i}(\phi)=M_i+y_i\phi .
\end{equation}
It is important to distinguish the physical heavy-neutrino mass
splitting from the parameters entering the effective potential. The radiative bias is determined by the difference of the effective
potential evaluated at the two vacua,
\begin{equation}
\Delta V_{\rm bias}
=
\left|
V_{\rm eff}(+v_\phi)-V_{\rm eff}(-v_\phi)
\right|,
\label{eq:bias_general}
\end{equation}
and therefore depends, in general, on the individual combinations
$M_i$ and $y_i$, rather than only on their particular combination
appearing in $\Delta M$. This distinction is essential when scanning
the quasi-degenerate regime, since different microscopic parameter
choices can yield the same $\Delta M$ while producing different
radiative vacuum-energy biases. The effective potential entering the numerical analysis is understood
to contain the contributions required by the adopted benchmark
prescription. When the annihilation temperature approaches the
symmetry-breaking or electroweak scale, finite-temperature corrections
should be included consistently. The radiative bias therefore depends on the individual combinations
$M_i$ and $y_i$ entering $m_{N_i}(\phi)=M_i+y_i\phi$. In particular,
different microscopic choices of $M_i$ and $y_i$ can yield the same
physical $\Delta M$ while producing different values of
$\Delta V_{\rm bias}$. Hence, the domain-wall bias must be calculated
directly from the effective potential for each microscopic parameter
point rather than inferred from $\Delta M$ alone.

We use the same zero-temperature wall tension,
\begin{equation}
\sigma_{\rm DW}=
\frac{2\sqrt{2\lambda_\phi}}{3}\,v_\phi^3,
\label{eq:sigma}
\end{equation}
throughout the numerical analysis.
The evolution of biased domain walls and their gravitational-wave
production have been studied extensively
\cite{Hiramatsu2013,Saikawa2017,Caprini2019}.

The competition between the bias pressure and the wall tension provides
the conventional analytic estimate for the onset of wall annihilation,
\begin{equation}
p_{\rm bias}\sim\Delta V_{\rm bias},
\qquad
p_{\rm DW}\sim\sigma_{\rm DW}H,
\label{eq:ann-condition}
\end{equation}
leading to
\begin{equation}
\Delta V_{\rm bias}(T_{\rm ann})
\sim
\sigma_{\rm DW}H(T_{\rm ann}).
\label{eq:Tann}
\end{equation}
We use Eq.~\eqref{eq:Tann} as the analytic criterion defining the
annihilation scale in the benchmark domain-wall treatment adopted
here. Recent numerical studies indicate that the nonlinear collapse
stage can modify the detailed spectral shape and normalization relative
to simple analytic estimates
\cite{Notari2025,Babichev2025}. We therefore interpret the resulting
GW spectrum within the domain-wall prescription specified explicitly
in the numerical analysis.

The annihilating wall network sources a stochastic gravitational-wave
background. We characterize the resulting spectrum by its peak
frequency and peak amplitude,
\begin{equation}
f_{\rm peak}=f_{\rm peak}(T_{\rm ann}),
\qquad
\Omega_{\rm GW}^{\rm peak}h^2
=
\Omega_{\rm GW}^{\rm peak}
(\sigma_{\rm DW},T_{\rm ann},g_*).
\label{eq:gw-characteristics}
\end{equation}
For the benchmark domain-wall prescription adopted in our analysis,
the redshifted peak frequency is approximately
\begin{equation}
f_{\rm peak}
\simeq
7.5\times10^{-9}\,{\rm Hz}
\left(\frac{T_{\rm GW}}{0.1\,{\rm GeV}}\right)
\left(\frac{g_*}{10}\right)^{1/6},
\label{eq:fpeak}
\end{equation}
where
\begin{equation}
T_{\rm GW}\simeq0.3\,T_{\rm ann}.
\end{equation}
The corresponding peak amplitude is determined by the wall tension,
annihilation temperature and relativistic degrees of freedom. In the
numerical analysis we use the complete expression associated with the
adopted domain-wall prescription rather than relying solely on the
scaling relation in Eq.~\eqref{eq:fpeak}.

The heavy-neutrino parameters play a dual role in the leptogenesis
and domain-wall sectors. The physical mass splitting
\begin{equation}
\Delta M =
(M_2-M_1)-(y_2-y_1)v_\phi
\label{eq:physical-splitting}
\end{equation}
controls the degree of heavy-neutrino degeneracy and hence the
quantum-kinetic evolution. At the same time, the individual parameters
$M_i$ and $y_i$ enter the scalar effective potential and therefore the
radiatively generated vacuum-energy bias. Thus, in general,
$\Delta V_{\rm bias}$ is not a function of $\Delta M$ alone. The
microscopic dependence is more appropriately represented as
\begin{equation}
\{M_1,M_2,y_1,y_2,v_\phi,\lambda_\phi\}
\longrightarrow
\{\Delta M,\Delta V_{\rm bias}\},
\label{eq:microscopic-map}
\end{equation}
while the subsequent domain-wall evolution gives
\begin{equation}
\Delta V_{\rm bias}
\longrightarrow
T_{\rm ann}
\longrightarrow
\left(
f_{\rm peak},
\Omega_{\rm GW}^{\rm peak}h^2
\right).
\label{eq:DW-map}
\end{equation}
In the leptogenesis sector, the same physical splitting determines
the degree of heavy-neutrino overlap through
\begin{equation}
\Delta M
\longrightarrow
\frac{\Delta M}{\Gamma_{N_1}}
\longrightarrow
Y_B^{\rm QKE}.
\label{eq:QKE-chain}
\end{equation}
Consequently, the central result of this work is a
model-dependent correlation between the quantum-kinetic leptogenesis
and domain-wall gravitational-wave sectors, rather than a one-to-one
mapping between $\Delta M$ and the gravitational-wave observables.

The existence of a seesaw-assisted domain-wall gravitational-wave
signal is not itself claimed as a new result, since it has been
established previously~\cite{KitajimaLeeMuraiTakahashiYin2024}. The distinctive feature
of the present analysis is instead the use of
$\Delta M/\Gamma_{N_1}$ to resolve the separated, resonant and
quantum-coherent heavy-neutrino regimes and to map these regimes onto
the predicted gravitational-wave parameter space. This provides the
basis for the quantum-kinetic--gravitational-wave correlation studied
in the following sections.

\section{Quantum-kinetic resonant leptogenesis}
\label{sec:lepto}

For quasi-degenerate heavy neutrinos, the mass splitting can become
comparable to the interaction-induced damping scale. In this regime the
off-diagonal entries of the heavy-neutrino density matrix encode coherent
oscillations and cannot in general be neglected
\cite{Pilaftsis1997,PilaftsisUnderwood2004,
GarnyKartavtsevHohenegger2013,Dev2014,Jukkala2021}.

As a classical reference, we use the conventional Boltzmann system
\begin{align}
 \frac{\dd Y_{N_i}}{\dd z}
 &=
 -\frac{1}{sHz}
 \left[
 \gamma_{D_i}
 \left(\frac{Y_{N_i}}{Y_{N_i}^{\rm eq}}-1\right)+\cdots
 \right],
 \nonumber\\
 \frac{\dd Y_{B-L}}{\dd z}
 &=
 \frac{1}{sHz}
 \left[
 \sum_i\epsilon_i\gamma_{D_i}
 \left(\frac{Y_{N_i}}{Y_{N_i}^{\rm eq}}-1\right)
 -W\,Y_{B-L}
 \right],
 \label{eq:boltzmann}
\end{align}
where $z=M_{N_1}/T$, $s$ is the entropy density, $H$ is the Hubble
rate, and $\gamma_{D_i}$ denotes the thermally averaged decay reaction
density. The final baryon asymmetry is
\begin{equation}
Y_B=\frac{28}{79}Y_{B-L}.
\label{eq:sphaleron}
\end{equation}
Equation~\eqref{eq:boltzmann} is used only as a classical reference for
the quantum-kinetic calculation.

The heavy-neutrino density matrix is
\begin{equation}
\rho_N=
\begin{pmatrix}
\rho_{11}&\rho_{12}\\
\rho_{21}&\rho_{22}
\end{pmatrix}.
\end{equation}
We evolve it with the flavour-covariant density-matrix transport
prescription
\begin{equation}
\frac{\dd\rho_N}{\dd z}
=
-\frac{i}{Hz}[H_N,\rho_N]
-\frac{1}{2Hz}
\left\{\Gamma_N,\rho_N-\rho_N^{\rm eq}\right\},
\label{eq:qke}
\end{equation}
with
\begin{equation}
H_N=\frac{M_N^2}{2E}+V_N(T).
\label{eq:HN}
\end{equation}
Here $\Gamma_N$ is the interaction-induced damping matrix and
$V_N(T)$ contains the thermal dispersive contribution. The scattering
and CP-violating source terms are included through the flavour-covariant
transport kernels described in Refs.~\cite{BenekeGarbrechtFidlerHerranenSchwaller2011,
DevMillingtonPilaftsisTeresi2014,JukkalaKainulainenRahkila2021}.
For clarity, the corresponding lepton-flavour asymmetry matrix
$\Delta_\ell$ is evolved in compact form as
\begin{equation}
\frac{\dd\Delta_\ell}{\dd z}
=
\frac{1}{Hz}
\left[
{\cal S}_\ell[\rho_N]
-\frac12\{{\cal W}_\ell,\Delta_\ell\}
-i[\Lambda_\ell,\Delta_\ell]
\right],
\label{eq:flavour-qke}
\end{equation}
where ${\cal S}_\ell$ is the CP-violating source, ${\cal W}_\ell$ the
washout matrix, and $\Lambda_\ell$ the thermal flavour-dispersion
matrix. Their entries are constructed from the Yukawa couplings and
thermal reaction densities according to the flavour-covariant
transport formalism cited above. The baryon asymmetry is obtained from
the trace of the final $B-L$ density after sphaleron conversion. This
form makes explicit that the $\delta_{\rm CP}$ dependence, when present,
arises through flavour-dependent source and washout terms rather than
from an unflavoured combination $h^\dagger h$.

We take vanishing initial lepton asymmetry,
$\Delta_\ell(z_{\rm in})=0$, and an initially thermal heavy-neutrino
density matrix,
$\rho_N(z_{\rm in})=\rho_N^{\rm eq}\mathbb I$, unless otherwise stated.
The system is evolved to a temperature well below the heavy-neutrino
mass, where the asymmetry has reached its asymptotic value. The
resonant regulator and thermal rates are those of the adopted
flavour-covariant transport prescription; no separate resonant
``$\epsilon_i$'' formula is used to replace the density-matrix source.

For organizing the numerical scan we define the vacuum-width ratio
\begin{equation}
r_q\equiv\frac{\Delta M}{\Gamma_{N_1}},
\label{eq:rq-qke}
\end{equation}
with
\begin{equation}
\Gamma_{N_1}
=
\frac{(h^\dagger h)_{11}M_{N_1}}{8\pi}.
\label{eq:width}
\end{equation}
The ratio $r_q$ is used as a diagnostic coordinate rather than as a
sharp phase boundary; the actual kinetic evolution is determined by
the temperature-dependent rates in Eq.~\eqref{eq:qke}. We distinguish
\begin{equation}
\begin{gathered}
r_q\gg1:\ {\rm separated},\qquad
r_q\sim1:\ {\rm resonant},\\
r_q\ll1:\ {\rm strongly\ coherent}.
\end{gathered}
\end{equation}

A useful numerical diagnostic is
\begin{equation}
{\cal R}_B\equiv
\frac{Y_B^{\rm QKE}}{Y_B^{\rm Boltzmann}}.
\label{eq:ratio}
\end{equation}
In the limit in which coherent effects and the associated off-diagonal
transport become negligible, the two calculations should approach one
another, providing the consistency check
\begin{equation}
r_q\gg1\quad\Longrightarrow\quad{\cal R}_B\to1.
\end{equation}
The principal quantity of interest is therefore the correlation between
$r_q$, the quantum-kinetic baryon asymmetry, and the domain-wall GW
observables.

\section{Numerical analysis and results}
\label{sec:num}

We take normal ordering as the reference realization of the minimal
two-right-handed-neutrino seesaw. Since the rank of the light-neutrino
mass matrix is two, one light neutrino is massless at tree level; hence
we do not scan $m_{\rm lightest}$ independently. In the numerical analysis, the physical heavy-neutrino splitting
$\Delta M=M_{N_2}-M_{N_1}$ is used as a scan coordinate, whereas
the corresponding field-independent mass parameters $M_i$ are
reconstructed consistently from the physical masses and the
scalar--right-handed-neutrino couplings $y_i$. The vacuum-energy
bias $\Delta V_{\rm bias}$ is then calculated independently from
the field-dependent masses entering the effective potential.
This avoids identifying the physical mass splitting with the
radiative vacuum-energy bias. The resulting gravitational-wave
prediction is consequently a correlated projection of the underlying
microscopic parameter space. For normal ordering,
\begin{equation}
m_1=0,\qquad
m_2=\sqrt{\Delta m^2_{21}},\qquad
m_3=\sqrt{\Delta m^2_{31}}.
\label{eq:NO-masses}
\end{equation}
The oscillation parameters are taken from NuFIT 6.0
\cite{NuFIT2024}. For the reference numerical configuration we use
\begin{equation}
\begin{gathered}
\sin^2\theta_{12}=0.308,\qquad
\sin^2\theta_{23}=0.470,\qquad
\sin^2\theta_{13}=0.02215,\\
\Delta m^2_{21}=7.49\times10^{-5}\ {\rm eV}^2,\qquad
\Delta m^2_{31}=2.513\times10^{-3}\ {\rm eV}^2,
\end{gathered}
\label{eq:nufit-values}
\end{equation}
while $\delta_{\rm CP}$ is varied explicitly in the dedicated scan.
The illustrative value $\delta_{\rm CP}=-90^\circ$ used for some
benchmark plots is therefore a scan point and is not identified with
the current global best fit.

The Casas--Ibarra matrix is
\begin{equation}
h =
\frac{i\sqrt{2}}{v}\,
U_{\rm PMNS}\sqrt{m_\nu^{\rm diag}}\,
R^\dagger\sqrt{M_N},
\qquad
RR^T=\mathbb I_{2\times2},
\label{eq:CI-num}
\end{equation}
with the normal-ordering form
\begin{equation}
R_{\rm NO}=
\begin{pmatrix}
0&\cos z_{\rm CI}&\sin z_{\rm CI}\\
0&-\sin z_{\rm CI}&\cos z_{\rm CI}
\end{pmatrix},
\qquad
z_{\rm CI}=x+iy .
\end{equation}
The factor of $i$ in Eq.~\eqref{eq:CI-num} is required by the sign
convention in the seesaw relation
$M_\nu=-M_D M_N^{-1}M_D^T$.

Unless otherwise stated, the numerical analysis is performed around
the reference parameter point
\begin{equation}
M_{N_1}=10^7~{\rm GeV},\qquad
v_\phi=10^6~{\rm GeV},\qquad
z_{\rm CI}=0.28+0.15i,\qquad
\lambda_\phi=0.1.
\label{eq:reference_point}
\end{equation}

For numerical reproducibility, the complete set of benchmark
parameters used in the reference calculation is summarized in
Table~\ref{tab:benchmark}. In particular, the scalar--right-handed-neutrino
couplings $y_1$ and $y_2$ are fixed to the explicit reference values
shown in the table. The heavy-neutrino mass splitting, the
Casas--Ibarra parameters, and the low-energy CP phase are varied
according to the scan prescription described below.

\begin{table}[t]
\centering
\caption{Reference benchmark parameters and scan ranges used in the
numerical analysis. Parameters not varied in a particular projection
are fixed to their reference values.}
\label{tab:benchmark}
\begin{tabular}{lcc}
\hline\hline
Parameter & Reference value & Scan range \\
\hline
$M_{N_1}$ &
$10^{7}\,{\rm GeV}$ &
$10^{5}$--$10^{9}\,{\rm GeV}$ \\

$\Delta M$ &
$10^{-3}\,{\rm GeV}$ &
$10^{-6}$--$10^{2}\,{\rm GeV}$ \\

$v_\phi$ &
$10^{6}\,{\rm GeV}$ &
$10^{5}$--$10^{9}\,{\rm GeV}$ \\

$\lambda_\phi$ &
$0.1$ &
fixed \\

$y_1$ &
$1.0\times10^{-3}$ &
fixed \\

$y_2$ &
$1.1\times10^{-3}$ &
fixed \\

$\mathrm{Re}(z_{\rm CI})$ &
$0.28$ &
fixed \\

$\mathrm{Im}(z_{\rm CI})$ &
$0.15$ &
fixed \\

$\delta_{\rm CP}$ &
$-90^\circ$ &
$-180^\circ$ \;--\; $180^\circ$ \\
\hline\hline
\end{tabular}
\end{table}
The scalar--right-handed-neutrino couplings are fixed to
$y_1=1.0\times10^{-3}$ and $y_2=1.1\times10^{-3}$ throughout
the reference analysis. These small non-degenerate couplings
provide the explicit $Z_2$-breaking source responsible for the
radiatively generated vacuum-energy bias while preserving the
quasi-degenerate heavy-neutrino configuration.
The parameter space is explored by logarithmic sampling in
$M_{N_1}$, $\Delta M$, and $v_\phi$ over the ranges shown in
Table~\ref{tab:benchmark}, while $\delta_{\rm CP}$ is sampled
uniformly over $[-180^\circ,180^\circ]$. Unless explicitly varied,
all remaining parameters are fixed to their reference values. For every sampled
microscopic parameter point, the physical heavy-neutrino masses,
the Yukawa couplings obtained from the Casas--Ibarra
parametrization, the decay width $\Gamma_{N_1}$, and the
radiatively generated vacuum-energy bias $\Delta V_{\rm bias}$ are
calculated consistently.
The two heavy neutrinos are taken to be quasi-degenerate. For each
parameter point, the physical masses are parametrized as
\begin{equation}
M_{N_2}=M_{N_1}+\Delta M,
\label{eq:MN2}
\end{equation}
with
\begin{equation}
|\Delta M|\ll M_{N_1}.
\label{eq:quasidegenerate}
\end{equation}

The scalar--right-handed-neutrino couplings are fixed to the
reference values
\begin{equation}
y_1=y_1^{\rm ref},\qquad
y_2=y_2^{\rm ref},
\end{equation}
where the explicit numerical values of $y_1^{\rm ref}$ and
$y_2^{\rm ref}$ are given in Table~\ref{tab:benchmark}. This makes
the numerical prescription fully specified and allows the
field-independent mass parameters to be reconstructed as
\begin{equation}
M_i=M_{N_i}+y_i v_\phi.
\label{eq:reconstruct_Mi}
\end{equation}
The field-dependent masses entering the scalar effective potential
are therefore
\begin{equation}
m_{N_i}(\phi)=M_i+y_i\phi.
\label{eq:fielddependent_mass}
\end{equation}

Consequently, the physical mass splitting satisfies
\begin{equation}
\Delta M=(M_2-M_1)-(y_2-y_1)v_\phi
=M_{N_2}-M_{N_1}.
\end{equation}
For each parameter point, the Yukawa couplings are reconstructed from
the Casas--Ibarra parametrization using the physical heavy-neutrino
masses $M_{N_i}$. The decay width is then calculated as
\begin{equation}
\Gamma_{N_1}
=
\frac{(h^\dagger h)_{11}M_{N_1}}{8\pi},
\end{equation}
and the density-matrix system of Sec.~\ref{sec:lepto} is solved to
obtain $Y_B^{\rm QKE}$. Independently, the same parameter point is
propagated through the scalar effective potential to calculate
$\Delta V_{\rm bias}$, $T_{\rm ann}$, and the gravitational-wave
observables. Thus, $\Delta V_{\rm bias}$ is evaluated directly from
the field-dependent heavy-neutrino masses and is not inferred from
$\Delta M$ alone.
For each parameter point, the Yukawa couplings are reconstructed from
the Casas--Ibarra parametrization using the physical heavy-neutrino
masses $M_{N_i}$. We then calculate the vacuum decay width
\begin{equation}
\Gamma_{N_1}
=
\frac{(h^\dagger h)_{11}M_{N_1}}{8\pi},
\end{equation}
and solve the density-matrix system of Sec.~\ref{sec:lepto} to obtain
$Y_B^{\rm QKE}$. The same microscopic parameter point is subsequently
propagated through the scalar effective potential to determine
$\Delta V_{\rm bias}$, $T_{\rm ann}$, and the corresponding
gravitational-wave observables. Thus, the numerical calculation
consistently projects the same microscopic parameter space onto both
the quantum-kinetic leptogenesis and domain-wall gravitational-wave
sectors.

The lower-scale point $M_{N_1}=500$ GeV with $\Delta M=1$ keV is
retained only as a reference for comparison with the published
seesaw-assisted domain-wall study of Ref.~\cite{KitajimaLeeMuraiTakahashiYin2024};
it is not used to define the main high-scale scan.
\begin{equation}
\boxed{
\begin{aligned}
\{\delta_{\rm CP},M_{N_1},v_\phi,\Delta M,z_{\rm CI},y_1,y_2,\lambda_\phi\}
&\longrightarrow
h_{\alpha i}
\longrightarrow
\left(r_q,Y_B^{\rm QKE}\right),
\\[2mm]
\{\delta_{\rm CP},M_{N_1},v_\phi,\Delta M,z_{\rm CI},y_1,y_2,\lambda_\phi\}
&\longrightarrow
\Delta V_{\rm bias}
\longrightarrow
T_{\rm ann}
\longrightarrow
\left(f_{\rm peak},\Omega_{\rm GW}^{\rm peak}h^2\right).
\end{aligned}}
\label{eq:numerical-map}
\end{equation}
The scan is divided into four one-dimensional projections:
\begin{table}[t]
\centering
\begin{tabular}{lll}
\toprule
Scan & Fixed parameters & Varied parameter\\
\midrule
A &
$M_{N_1},v_\phi,z_{\rm CI}$ & $\delta_{\rm CP}$\\
B &
$v_\phi,z_{\rm CI},\delta_{\rm CP}$ & $M_{N_1}$\\
C &
$M_{N_1},z_{\rm CI},\delta_{\rm CP}$ & $v_\phi$\\
D &
$M_{N_1},v_\phi,z_{\rm CI},\delta_{\rm CP}$ & $\Delta M$\\
\bottomrule
\end{tabular}
\caption{Reference projections used to expose the dependence of the
quantum-kinetic and gravitational-wave observables on the microscopic
parameters. The lightest neutrino mass is fixed to zero, as required by
the minimal two-RHN construction.}
\label{tab:scans}
\end{table}

A point is retained as phenomenologically viable only when all
relevant requirements are satisfied simultaneously:
\begin{enumerate}
\item neutrino oscillation constraints;
\item perturbativity and validity of the seesaw expansion;
\item the flavour-sensitive quantum-kinetic evolution;
\item agreement with the observed baryon asymmetry,
\begin{equation}
Y_B^{\rm obs}\simeq8.7\times10^{-11};
\end{equation}
\item formation and annihilation of the domain-wall network without
    unacceptable domination;
\item BBN/CMB bounds on the integrated GW energy density; and
\item consistency with the adopted GW sensitivity and cosmological
    bounds.
\end{enumerate}
For the numerical selection we use
\begin{equation}
\left|
\frac{Y_B^{\rm QKE}}{Y_B^{\rm obs}}-1
\right|<\epsilon_{\rm BAU},
\label{eq:BAU-selection}
\end{equation}
where $\epsilon_{\rm BAU}$ denotes the numerical tolerance used in the
scan.

\subsection{Quantum-kinetic--gravitational-wave correlation}
\label{subsec:qke-gw-correlation}

The central result is the correlation between the quantum regime of the
quasi-degenerate RHN system and the primordial GW signal. We characterize
the former by $r_q=\Delta M/\Gamma_{N_1}$ and quantify the difference
from the classical treatment through
\begin{equation}
{\cal R}_B=
\frac{Y_B^{\rm QKE}}{Y_B^{\rm Boltzmann}}.
\end{equation}
For large $r_q$, the off-diagonal heavy-neutrino correlations become
progressively less important and the density-matrix result should
approach the classical limit, ${\cal R}_B\to1$. A departure from unity
therefore provides a diagnostic of quantum-coherent effects.

\begin{figure*}[!t]
\centering
\includegraphics[width=0.98\textwidth]{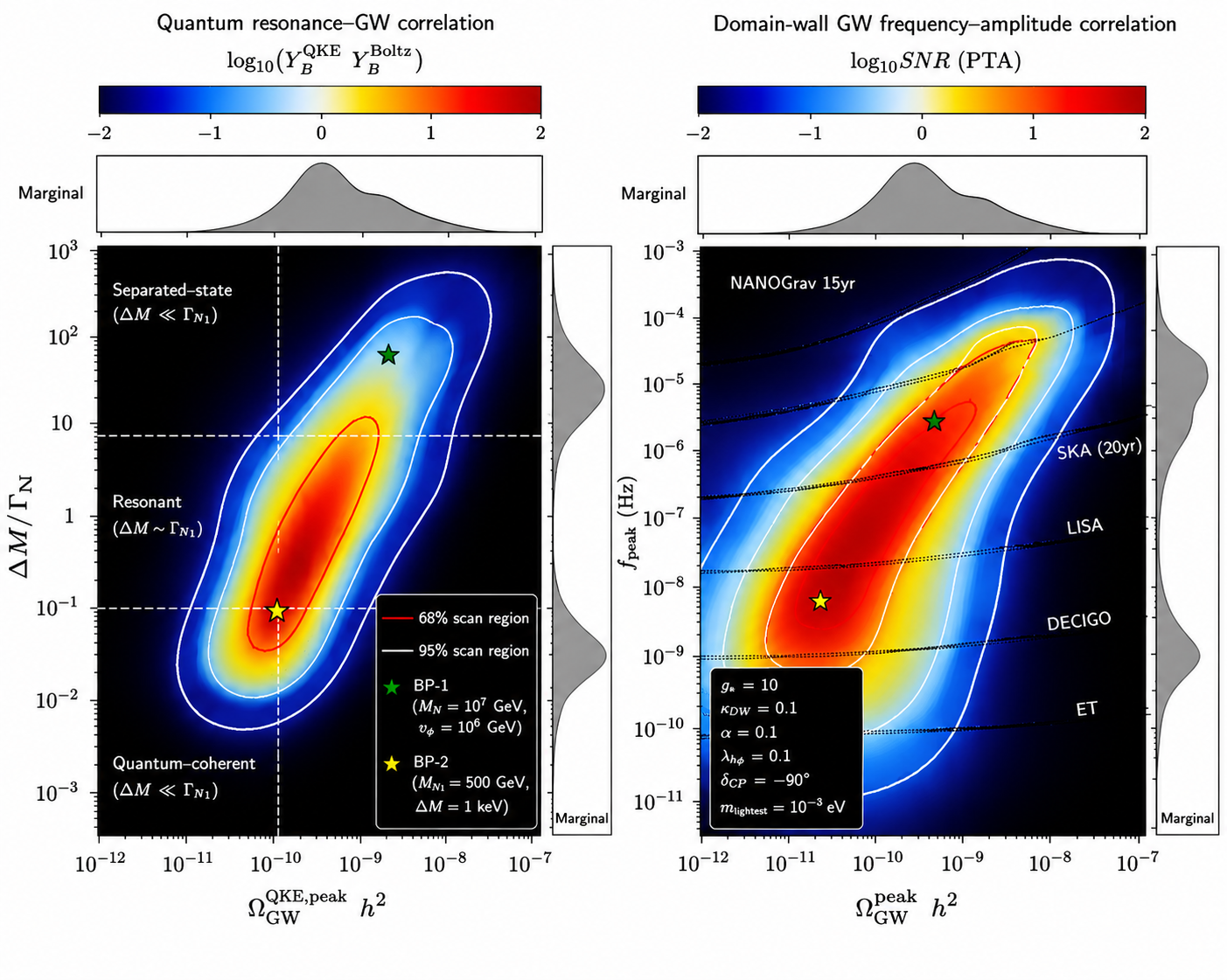}
\caption{Quantum-kinetic--gravitational-wave parameter-space correlation
in the seesaw-assisted domain-wall framework. The left panel shows the
microscopic scan projected onto the
$\log_{10}(\Delta M/\Gamma_{N_1})$--$\log_{10}
(\Omega_{\rm GW}^{\rm peak}h^2)$ plane; the colour scale represents
$\log_{10}(Y_B^{\rm QKE}/Y_B^{\rm Boltzmann})$. The right panel shows
the corresponding projection onto the observable
$(f_{\rm peak},\Omega_{\rm GW}^{\rm peak}h^2)$ plane together with
representative gravitational-wave sensitivities. The quantity
$\Delta M/\Gamma_{N_1}$ characterizes the quantum-kinetic regime of
the quasi-degenerate heavy-neutrino system, while the domain-wall
gravitational-wave observables are obtained from the radiatively
generated vacuum-energy bias calculated from the underlying
field-dependent heavy-neutrino masses
$m_{N_i}(\phi)=M_i+y_i\phi$. Thus, the two panels represent correlated
projections of the same microscopic parameter space rather than a
one-to-one mapping from $\Delta M/\Gamma_{N_1}$ to the gravitational-wave
signal. The quantum-kinetic treatment follows the density-matrix and
flavour-covariant resonant-leptogenesis literature
~\cite{Pilaftsis1997,PilaftsisUnderwood2004,GarnyKartavtsevHohenegger2013,
Dev2014,Jukkala2021,BenekeGarbrechtFidlerHerranenSchwaller2011,
DevMillingtonPilaftsisTeresi2014,JukkalaKainulainenRahkila2021,LiPilaftsis2026},
while the domain-wall gravitational-wave interpretation follows the
standard annihilation literature and recent numerical studies
~\cite{Hiramatsu2013,Saikawa2017,Notari2025,KitajimaLeeMuraiTakahashiYin2024,
Babichev2025,NotariRompineveTorrenti2025,CyrCotterillBattye2025}.
The vertical divisions indicate separated, resonant, and strongly
coherent regimes only as an organizing classification in
$r_q=\Delta M/\Gamma_{N_1}$; they are not sharp physical phase
boundaries. The contours represent scan densities rather than
confidence or posterior-probability regions, and the benchmark markers
denote reference configurations rather than automatically
phenomenologically allowed points.}
\label{fig:QKE-GW-correlation}
\end{figure*}

The same microscopic parameter points are propagated through the
domain-wall calculation. The radiative vacuum-energy bias is
calculated directly from the field-dependent heavy-neutrino masses,
\begin{equation}
m_{N_i}(\phi)=M_i+y_i\phi,
\end{equation}
and therefore depends on the underlying parameters $M_i$ and $y_i$.
The domain-wall evolution subsequently gives
\begin{equation}
\Delta V_{\rm bias}
\longrightarrow
T_{\rm ann}
\longrightarrow
\left(
f_{\rm peak},
\Omega_{\rm GW}^{\rm peak}h^2
\right).
\label{eq:GW-chain-corrected}
\end{equation}
The quantum-kinetic coordinate
$r_q=\Delta M/\Gamma_{N_1}$ can therefore be used to organize the
resulting parameter-space projection onto the observable
gravitational-wave plane, but it should not be interpreted as uniquely
fixing the domain-wall bias or the gravitational-wave signal.
Figure~\ref{fig:QKE-GW-correlation} displays this correlated
multi-parameter projection. Figure~\ref{fig:QKE-GW-correlation} displays this mapping.

The quantum-kinetic correlation is complemented by the neutrino-sector
dependence of the predicted GW background.  We retain the corresponding
multi-panel result from the original numerical analysis as an auxiliary
sensitivity study.  In particular, the dependence on $\delta_{\rm CP}$
is interpreted only within the flavour-sensitive kinetic treatment.
The panel showing a variation of $m_{\rm lightest}$ should not be interpreted as part of the strict minimal two-RHN tree-level scan, for which one light neutrino is massless. It is retained only as an auxiliary comparison with a broader Casas--Ibarra parameterization; no point from that panel is used in the combined viable-region selection.

\begin{figure*}[!b]
 \centering
 \includegraphics[width=0.98\textwidth]{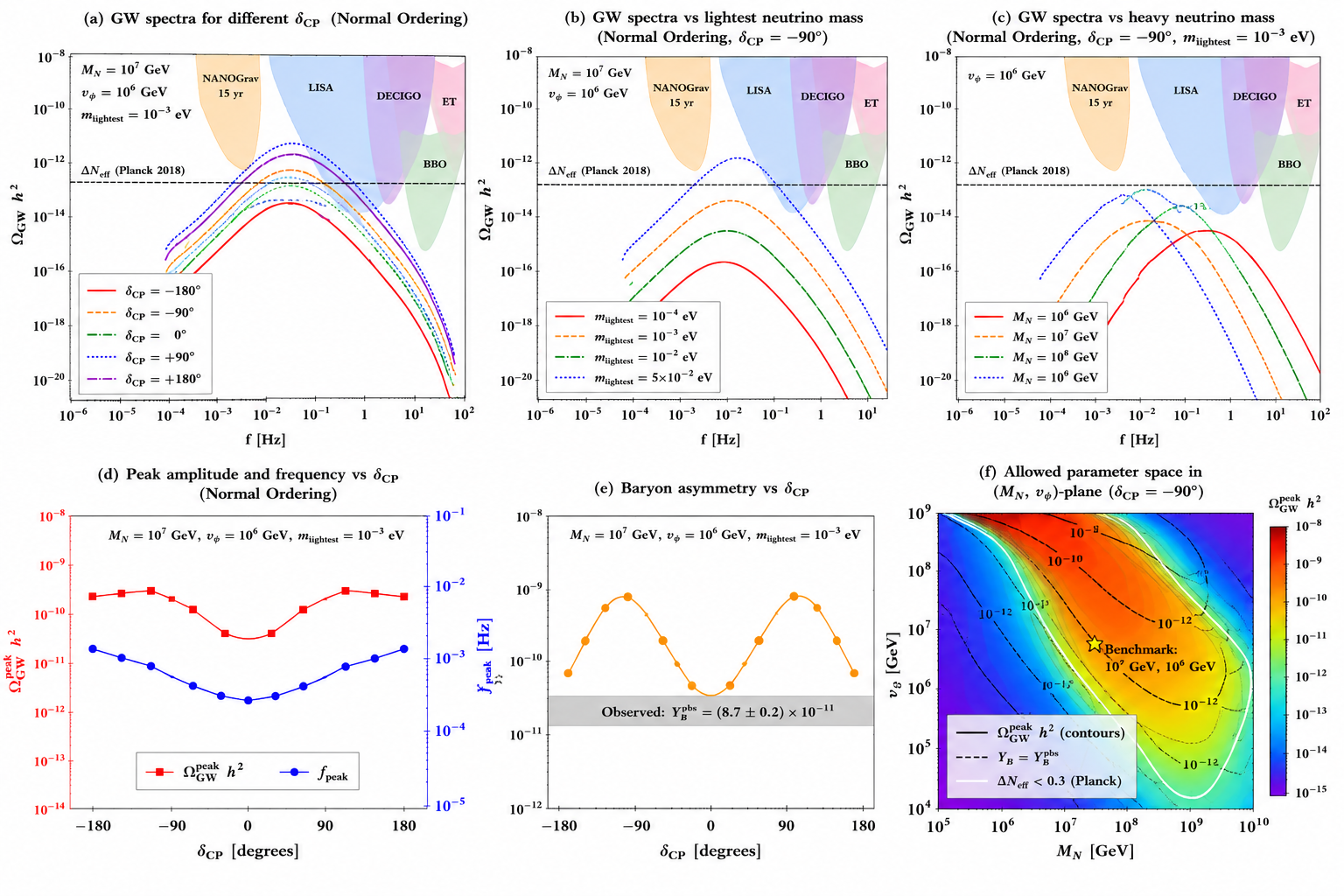}
 \caption{\label{fig:neutrino-gw-main}
 Neutrino-sector dependence of the primordial gravitational-wave
 background in the seesaw-assisted domain-wall framework. Panel (a)
 shows the GW spectra for different values of the Dirac CP phase
 $\delta_{\rm CP}$ at fixed heavy-neutrino and scalar scales. Panel (b)
 shows the sensitivity to the lightest-neutrino mass in the broader
 Casas--Ibarra scan. Panel (c) displays the dependence on the
 heavy-neutrino scale. Panel (d) shows the corresponding movement of the
 GW peak in the $(f_{\rm peak},\Omega_{\rm GW}^{\rm peak}h^2)$ plane.
 Panel (e) displays the baryon asymmetry obtained from the
 flavour-sensitive quantum-kinetic calculation as a function of
 $\delta_{\rm CP}$, while panel (f) shows the combined
 $(M_{N_1},v_\phi)$ parameter space after imposing the adopted
 neutrino, leptogenesis, domain-wall and cosmological requirements.
 The low-energy CP interpretation follows the flavour-dependent
 leptogenesis literature~\cite{BrancoMorozumiNobreRebelo2002,
 AbadaDavidLosada2006,PascoliPetcovRiotto2007,BenekeGarbrechtFidlerHerranenSchwaller2011,
 DevMillingtonPilaftsisTeresi2014,JukkalaKainulainenRahkila2021},
 while the neutrino inputs and absolute-mass constraints are connected to
 current oscillation and cosmological analyses~\cite{NuFIT2024,Gariazzo2024,
 DiValentino2024,AllaliNotari2024,CasasIbarra2001}. The contours represent
 parameter-scan correlations and are not posterior probability or confidence
 regions. The $m_{\rm lightest}$ panel is presented as an auxiliary sensitivity
 study and is not used to claim a second independent nonzero light-neutrino
 mass in the strict minimal two-RHN tree-level limit.}
\end{figure*}

As shown in Fig.~\ref{fig:neutrino-gw-main}, the dependence on the low-energy neutrino sector is indirect and is mediated by the Casas--Ibarra Yukawa couplings, the RHN interaction rates and the radiative domain-wall bias. The right-hand panel should therefore be interpreted as a correlated
projection rather than as a direct measurement of $\Delta M$ or any
individual low-energy neutrino parameter. In particular, the GW
background does not uniquely determine $\delta_{\rm CP}$; any dependence
on the low-energy phase arises indirectly through the flavour-sensitive
kinetic evolution and the Casas--Ibarra Yukawa couplings.

\subsection{Gravitational-wave spectrum and domain-wall prescription}
\label{subsec:GW-spectra}

We parameterize the spectrum as
\begin{equation}
\Omega_{\rm GW}h^2
=
\Omega_{\rm GW}^{\rm peak}h^2\,
S\!\left(\frac{f}{f_{\rm peak}}\right),
\end{equation}
where $S(1)=1$. For the baseline phenomenological representation we use
\begin{equation}
S(x)=
\frac{(a+b)^c}
{\left[b\,x^{-a/c}+a\,x^{b/c}\right]^c}.
\end{equation}
This is a compact representation of the adopted domain-wall spectrum,
not a universal prediction for every wall potential. Recent numerical
studies show that the nonlinear annihilation stage can modify the
normalization and spectral shape relative to simple scaling estimates
\cite{NotariRompineveTorrenti2025,CyrCotterillBattye2025}. We therefore
treat the analytic relations below as scaling relations and state the
domain-wall prescription explicitly when quoting numerical predictions.

The characteristic redshifted peak frequency is
\begin{equation}
f_{\rm peak}\simeq
7.5\times10^{-9}\ {\rm Hz}
\left(\frac{T_{\rm GW}}{0.1~{\rm GeV}}\right)
\left(\frac{g_*}{10}\right)^{1/6},
\qquad
T_{\rm GW}\simeq0.3\,T_{\rm ann}.
\label{eq:fpeak-results}
\end{equation}
The corresponding peak amplitude scales as
\begin{equation}
\Omega_{\rm GW}^{\rm peak}h^2
\simeq
10^{-10}
\left(\frac{0.1~{\rm GeV}}{T_{\rm GW}}\right)^4
\left(\frac{\sigma_{\rm DW}}{10^{15}~{\rm GeV}^3}\right)^2
\left(\frac{10}{g_*}\right)^{4/3},
\label{eq:OmegaGW-results}
\end{equation}
with the numerical coefficient understood to be prescription dependent.
The robust scaling is
\begin{equation}
f_{\rm peak}\propto T_{\rm GW}g_*^{1/6},
\qquad
\Omega_{\rm GW}^{\rm peak}h^2
\propto\sigma_{\rm DW}^2T_{\rm GW}^{-4}g_*^{-4/3}.
\end{equation}

\begin{figure*}[!t]
    \centering
    \includegraphics[width=0.98\textwidth]{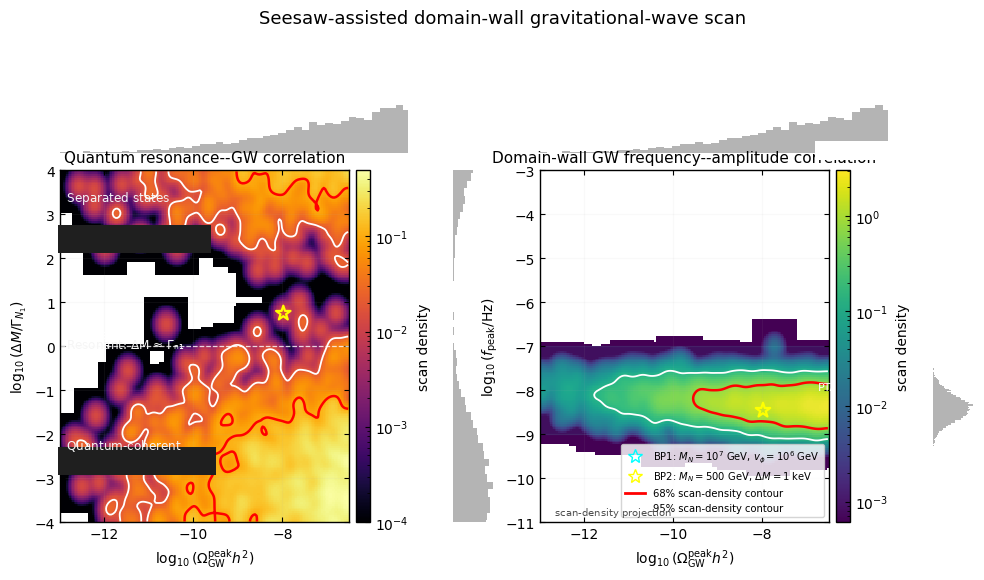}
    \caption{\label{fig:combined-qke-gw}
    Correlation between the quantum-kinetic regime of the quasi-degenerate
    right-handed-neutrino system and the primordial gravitational-wave
    signal in the seesaw-assisted domain-wall scenario. The left panel
    shows the scan-density distribution in the
    $\log_{10}(\Delta M/\Gamma_{N_1})$--$\log_{10}
    (\Omega_{\rm GW}^{\rm peak}h^2)$ plane. The right panel shows the
    corresponding projection onto the
    $\log_{10}(f_{\rm peak}/{\rm Hz})$--$\log_{10}
    (\Omega_{\rm GW}^{\rm peak}h^2)$ plane. The physical ingredients
    entering the correlation are the resonant density-matrix dynamics
    ~\cite{Pilaftsis1997,PilaftsisUnderwood2004,GarnyKartavtsevHohenegger2013,
    Dev2014,Jukkala2021,LiPilaftsis2026} and the biased-domain-wall GW
    spectrum~\cite{Hiramatsu2013,Saikawa2017,Notari2025,
    KitajimaLeeMuraiTakahashiYin2024,Babichev2025,
    NotariRompineveTorrenti2025,CyrCotterillBattye2025}. Related
    neutrino--GW studies provide complementary probes of seesaw scales,
    leptogenesis and neutrino mass generation~\cite{Dror2020,
    DrorHiramatsuKohriMurayamaWhite2020,Blasi2020,BlasiBrdarSchmitz2020,
    Fornal2020,FornalShamsEsHaghi2020,BhandariDattaSil2024,
    Datta2025,GhoshalKohriNarendra2026,GhoshalKohriNarendra2025,
    GhoshalPal2026,Wang2026}. The quantum-coherent, resonant-overlapping
    and separated-state regimes are identified according to
    $\Delta M/\Gamma_{N_1}$ only as an organizing classification.
    The contours represent scan-density regions and are not
    posterior-probability or confidence-level contours. Benchmark markers
    indicate the reference configurations used in the numerical analysis
    and are not automatically phenomenologically allowed points.}
\end{figure*}

Figure~\ref{fig:combined-qke-gw} provides a compact representation of the
same central correlation in a form suitable for comparison with the
observable GW plane.  It should be read together with
Fig.~\ref{fig:QKE-GW-correlation}: the two figures emphasize the same
physical mapping using the two plotting prescriptions retained from the
original analysis.

\subsection{Novelty and comparison with previous work}
\label{subsec:novelty}

The generic connection between seesaw neutrinos, leptogenesis and
primordial gravitational waves is not new
\cite{Dror2020,Blasi2020,Fornal2020}. Likewise, stochastic GWs from
biased domain-wall annihilation are well established
\cite{Hiramatsu2013,Saikawa2017}. Recent lattice calculations further
show that the detailed wall spectrum can differ from simple analytic
scaling relations~\cite{NotariRompineveTorrenti2025,CyrCotterillBattye2025}.

The closest realization of the particle-physics setup is recent seesaw-assisted and domain-wall-collapse studies
\cite{KitajimaLeeMuraiTakahashiYin2024,BanerjeeYajnik2024}. That work establishes the RHN-induced radiative
bias and its correlation with the GW peak properties, and discusses
small RHN mass splittings relevant for resonant leptogenesis. The
present work does not repeat that result. Its narrower objective is to
propagate the quantum-coherent dynamics of a quasi-degenerate RHN pair
through a flavour-sensitive density-matrix treatment and correlate the
result with the same domain-wall GW sector.

The distinction can be summarized as
\begin{equation}
\frac{\Delta M}{\Gamma_{N_1}}
\longrightarrow
Y_B^{\rm QKE},
\qquad
\{M_i,y_i,v_\phi,\lambda_\phi\}
\longrightarrow
\Delta V_{\rm bias}
\longrightarrow
T_{\rm ann}
\longrightarrow
\left(
f_{\rm peak},\Omega_{\rm GW}^{\rm peak}h^2
\right).
\label{eq:full_correlation}
\end{equation}
The two sectors are correlated because they depend on the same
underlying heavy-neutrino and scalar parameters, but the physical
mass splitting $\Delta M$ does not uniquely determine
$\Delta V_{\rm bias}$. The resulting relation is therefore a
model-dependent parameter-space correlation rather than a one-to-one
mapping between $\Delta M$ and the gravitational-wave observables.
This is a model-dependent multi-parameter consistency relation, not a
claim that a GW measurement uniquely determines $\Delta M$,
$\delta_{\rm CP}$ or $m_{\rm lightest}$.

The role of flavour is essential. In an unflavoured treatment the
Casas--Ibarra combination $h^\dagger h$ is independent of the explicit
PMNS matrix because of unitarity. A physical dependence on low-energy
CP phases can therefore arise only after flavour-dependent source and
washout effects are retained
\cite{BrancoMorozumiNobreRebelo2002,AbadaDavidLosada2006,
PascoliPetcovRiotto2007,BenekeGarbrechtFidlerHerranenSchwaller2011,
DevMillingtonPilaftsisTeresi2014,JukkalaKainulainenRahkila2021}.
Our $\delta_{\rm CP}$ scans are interpreted in this flavour-sensitive
framework and not as predictions of the unflavoured Boltzmann equations.

The construction is also distinct from GW--leptogenesis scenarios based
on cosmic strings or first-order phase transitions
\cite{Athron2026,Datta2026,GhoshalPal2026,GhoshalKohriNarendra2026}.
Those scenarios involve different GW sources and do not contain the
radiatively biased $\mathbb Z_2$ domain-wall mechanism combined here with
quasi-degenerate RHN quantum kinetics.

\subsection{Consistency requirements}
\label{subsec:consistency}

The numerical interpretation requires simultaneous consistency of the
particle and cosmological sectors. In addition to the oscillation-data
constraints and the baryon-asymmetry condition in
Eq.~\eqref{eq:BAU-selection}, we require perturbative couplings and a
valid seesaw expansion, formation and annihilation of the wall network
before wall domination, and consistency with BBN, CMB and integrated-GW
bounds. Detector sensitivity curves in Fig.~\ref{fig:QKE-GW-correlation}
are used only to indicate prospective reach; they do not define
phenomenological exclusion regions.

The scan-density contours in the figures must likewise not be interpreted
as confidence or credible regions because no statistical likelihood is
constructed. A parameter point is called viable only when the complete
set of neutrino, leptogenesis, domain-wall and cosmological conditions is
satisfied.

\section{Conclusions}
\label{sec:conclusion}

We have developed a unified framework connecting the neutrino mass
generation mechanism, quantum-kinetic resonant leptogenesis and a
primordial stochastic gravitational-wave (GW) background through a
seesaw-assisted domain-wall sector. The setup contains two
quasi-degenerate right-handed neutrinos and a real scalar field charged
under a discrete $\mathbb{Z}_2$ symmetry. An approximately $\mathbb{Z}_2$-symmetric scalar potential allows a domain-wall
network to form, while the small right-handed-neutrino coupling provides
a radiative vacuum-energy bias that drives its annihilation. The resulting GW signal is therefore
controlled by the same microscopic parameters that govern the heavy
neutrino sector.

The main point of the analysis is that the RHN mass splitting,
\begin{equation}
r_q\equiv\frac{\Delta M}{\Gamma_{N_1}},
\end{equation}
plays a dual role. In the leptogenesis sector it determines the degree
of overlap and quantum coherence of the quasi-degenerate RHN states,
whereas in the scalar sector, when the bare heavy-neutrino masses are
also quasi-degenerate, the splitting is related to the RHN--scalar
couplings through
\begin{equation}
\Delta M=(M_2-M_1)-(y_2-y_1)v_\phi,
\qquad |\Delta M|\ll M_{N_1}.
\end{equation} Consequently, the same microscopic
quantity enters both the quantum-kinetic generation of the baryon
asymmetry and the radiatively induced domain-wall bias.

A density-matrix treatment is therefore essential in the region
$\Delta M\lesssim\Gamma_{N_1}$, where the two heavy-neutrino states
cannot be consistently regarded as independent classical species
\cite{Dev2014,Jukkala2021,JukkalaKainulainenRahkila2021}. The present framework explicitly
tracks this transition through $r_q$ and compares the resulting
quantum-kinetic baryon asymmetry with the conventional Boltzmann
description. The resulting quantity
\begin{equation}
\mathcal{R}_B
=
\frac{Y_B^{\rm QKE}}
     {Y_B^{\rm Boltzmann}}
\end{equation}
provides a direct measure of the quantum correction to the classical
leptogenesis prediction.

The principal phenomenological connection can instead be summarized
as two correlated projections of the same microscopic parameter space,
\begin{equation}
\boxed{
\begin{aligned}
\frac{\Delta M}{\Gamma_{N_1}}
&\longrightarrow
Y_B^{\rm QKE},
\\[2mm]
\{M_i,y_i,v_\phi,\lambda_\phi\}
&\longrightarrow
\Delta V_{\rm bias}
\longrightarrow
T_{\rm ann}
\longrightarrow
\left(
f_{\rm peak},
\Omega_{\rm GW}^{\rm peak}h^2
\right).
\end{aligned}}
\label{eq:conclusion-chain}
\end{equation}
The two projections are correlated because the same underlying
heavy-neutrino and scalar parameters enter both sectors. However,
$\Delta M$ alone does not uniquely determine the radiative
vacuum-energy bias. The gravitational-wave signal should therefore be
interpreted as a model-dependent multi-parameter consistency relation,
rather than as a direct measurement of the heavy-neutrino mass
splitting.
Thus, rather than claiming that the seesaw--domain-wall--GW connection
itself is new, we establish a quantitative framework in which the
quantum regime of resonant leptogenesis is correlated with the
gravitational-wave parameter space through the common underlying
heavy-neutrino and scalar parameters. This should be viewed as a
complementary probe of the otherwise inaccessible dynamics of
quasi-degenerate heavy Majorana neutrinos.

This result extends the seesaw-assisted domain-wall mechanism previously
studied in Ref.~\cite{KitajimaLeeMuraiTakahashiYin2024}. That work established the
RHN-induced radiative bias and its correlation with the GW signal,
whereas the present analysis adds the quantum-kinetic description of the
quasi-degenerate RHN system and uses $\Delta M/\Gamma_{N_1}$ as the
organizing variable connecting leptogenesis to the GW prediction.
Related connections between seesaw leptogenesis and gravitational waves
have also been explored in cosmic-string and phase-transition
scenarios~\cite{Dror2020,Blasi2020,GhoshalPal2026}, but these involve
different cosmological sources of the GW background.

The gravitational-wave prediction must ultimately be subjected to the
same consistency requirements as the leptogenesis calculation. In
particular, phenomenologically viable points must simultaneously
satisfy the neutrino-oscillation constraints, perturbativity and
seesaw validity, the flavour-sensitive quantum-kinetic evolution and
the observed baryon asymmetry,
\begin{equation}
Y_B^{\rm obs}\simeq 8.7\times10^{-11},
\end{equation}
as well as the cosmological requirements on domain-wall formation and
annihilation, BBN and CMB constraints, and the relevant GW bounds and
detector sensitivities. The parameter-space contours should therefore
be interpreted as scan correlations rather than probability or
confidence regions unless a statistical likelihood analysis is
performed.

In summary, the main outcome is the establishment of a
neutrino--leptogenesis--GW consistency relation,
\begin{equation}
\boxed{
\begin{aligned}
\text{neutrino data}
&\rightarrow \text{seesaw Yukawa couplings}\\
&\rightarrow \text{RHN quantum kinetics}\\
&\rightarrow \text{domain-wall annihilation}\\
&\rightarrow \text{primordial GW signal}.
\end{aligned}}
\end{equation}
If a parameter region simultaneously reproduces the observed neutrino
mass spectrum and baryon asymmetry and produces a GW signal within the
reach of present or future detectors, the stochastic background would
provide an indirect probe of the quantum dynamics of the heavy-neutrino
sector. Conversely, the absence of an allowed overlap would place
combined constraints on the RHN mass splitting, scalar
symmetry-breaking scale and seesaw parameters. The framework therefore
offers a testable connection between low-energy neutrino physics,
early-Universe leptogenesis and primordial gravitational waves.

\end{document}